\documentclass[a4paper]{article}

\usepackage{INTERSPEECH2022}
\usepackage{amsmath}
\usepackage{textcomp}
\usepackage{url}
\usepackage[utf8]{inputenc}
\usepackage{CJKutf8}
\usepackage{multirow}
\usepackage{comment}
\usepackage[colorlinks=true]{hyperref}

\newcommand{\jp}[1]{ \begin{CJK}{UTF8}{ipxm}#1\end{CJK} }

\ninept

\title{CALLS: Japanese Empathetic Dialogue Speech Corpus of \\ Complaint Handling and Attentive Listening in Customer Center}

\makeatletter
\def\name#1{\gdef\@name{#1\\}}
\makeatother
\name{{\em Yuki Saito$^{1}$, Eiji Iimori$^{1}$, Shinnosuke Takamichi$^1$, Kentaro Tachibana$^2$, and Hiroshi Saruwatari$^1$}}

\address{
$^1$The University of Tokyo, Japan, $^2$LINE Corp., Japan.
}
\email{
yuuki\_saito@ipc.i.u-tokyo.ac.jp
}

\begin{document}
\setlength{\abovedisplayskip}{5pt} % 式の上部のマージン
\setlength{\belowdisplayskip}{5pt} % 式の下部のマージン
\setlength\floatsep{6pt} % 図と図の間のマージン
\setlength\intextsep{6pt} % 図と本文の間のマージン
\setlength\textfloatsep{6pt} % 図と本文の間のマージン
\setlength{\dbltextfloatsep}{6pt} % ぶち抜きの図表と本文の間のマージン
\setlength{\dblfloatsep}{5pt}

\maketitle

\begin{abstract}
We present {\it CALLS}, a Japanese speech corpus that considers phone calls in a customer center as a new domain of empathetic spoken dialogue. The existing STUDIES corpus covers only empathetic dialogue between a teacher and student in a school. To extend the application range of empathetic dialogue speech synthesis (EDSS), we designed our corpus to include the same female speaker as the STUDIES teacher, acting as an operator in simulated phone calls. We describe a corpus construction methodology and analyze the recorded speech. We also conduct EDSS experiments using the CALLS and STUDIES corpora to investigate the effect of domain differences. The results show that mixing the two corpora during training causes biased improvements in the quality of synthetic speech due to the different degrees of expressiveness. Our project page of the corpus is \href{http://sython.org/Corpus/STUDIES-2}{here}.
\end{abstract}

\vspace{-5pt}
\section{Introduction}\label{sect:intro}
\vspace{-3pt}
% Background
Empathetic dialogue speech synthesis (EDSS)~\cite{saito22studies} is an emerging technology in the text-to-speech (TTS)~\cite{sagisaka88} research field. Its goal is to develop a friendly voice agent that can empathetically talk to humans with a speaking style suitable for the dialogue situation. Saito et al.~\cite{saito22studies} first constructed a Japanese speech corpus of empathetic dialogues named \textit{STUDIES}. This corpus considers chit-chat between a teacher and student in a school~\cite{warren15} as the empathetic dialogue situation and includes voices of a female actor who performs the teacher. Their baseline end-to-end EDSS model well reproduced the teacher's empathetic speaking styles by using the speaker's emotion label and chat history as the conditional features. Since EDSS is essential for the next-generation society where humans and robots will collaborate, developing speech corpora for various empathetic dialogue situations is vital for TTS research and its related fields, such as natural language processing and spoken dialogue systems.

\begin{figure}[t]
  \centering
  \includegraphics[width=0.8\linewidth, clip]{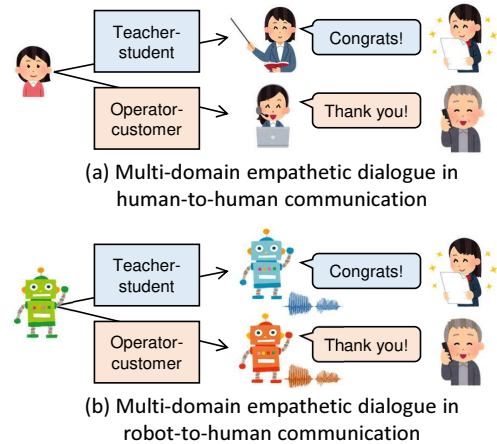}
  \vspace{-8pt}
  \caption{Multi-domain empathetic spoken dialogues in (a) human-to-human and (b) robot-to-human communications.}
  \label{fig:overview}
\end{figure}

Although humans can speak empathetically with their interlocutors in various dialogue domains, as shown in Fig.~\ref{fig:overview}(a), the STUDIES corpus assumes only a single dialogue domain. Concretely, the teacher attempts to motivate her students to increase their enjoyment of school life through her \textit{informal} (i.e., basically without honorific words) and intensely expressive speaking styles. However, such styles are only sometimes suitable for other dialogue domains where an empathetic listener talks more \textit{formally}, such as a doctor counseling a patient~\cite{hojat09}. Therefore, developing speech corpora containing a single speaker's empathetic dialogues in various domains will contribute to multi-domain EDSS that can appropriately control empathetic speaking styles in accordance with domains (Fig.~\ref{fig:overview}(b)).

To facilitate multi-domain EDSS research, we present {\it CALLS} (\textbf{C}omplaint handling and \textbf{A}ttentive \textbf{L}istening \textbf{L}ines \textbf{S}peech), a new empathetic dialogue speech corpus for developing EDSS technologies suitable for polite and formal dialogue domains. It includes the same female speaker as the STUDIES teacher, acting as a customer-center operator in simulated phone calls. This design enables a single speaker's different empathetic speaking styles to be analyzed in multiple dialogue domains. This paper presents 1) a methodology to construct our corpus, 2) the analysis results of the recorded speech, and 3) the results of EDSS experiments using the CALLS and STUDIES corpora to investigate the effect of domain differences. The contributions of this study are as follows:
    \vspace{-1mm}
    \begin{itemize} \leftskip -1mm \itemsep -0mm
        \item We construct a new corpus for developing a polite voice agent and building a multi-domain EDSS model. Our corpus is open-sourced for research purposes only from \href{http://sython.org/Corpus/STUDIES-2}{this page.}
        \item We analyze the dialogue domain differences between our CALLS and existing STUDIES corpora regarding the prosodic and textual features.
        \item We present the results of EDSS experiments and demonstrate that mixing the two corpora during the training causes biased improvements in the quality of synthetic speech due to the different degrees of expressiveness.
    \end{itemize}
    \vspace{-2mm}

\vspace{-5pt}
\section{Corpus Construction Methodology for Simulated Customer-Center Dialogues}
\vspace{-3pt}
\subsection{Dialogue scenario}\label{subsect:scenario}
\vspace{-3pt}
Empathetic behaviors by a customer-center operator can engender highly positive customer responses~\cite{clark13} and affect customer satisfaction~\cite{ando20}. However, recording such behaviors in actual operator-customer conversations suitable for TTS takes much work. The reason is that the contents may include much personal and confidential information, and the bandwidth is typically limited. Therefore, we instead considered simulated customer-center dialogues as the situation and constructed the CALLS corpus including two subsets: 1) situation-oriented complaint handling and 2) positive attentive listening.

{\bf Situation-oriented complaint handling:} The first subset focused on situations where customers feel dissatisfied with a product or service provided by a specific company or organization. The operator responds to the complaining customers by phone calls. We specified the operator's persona as ``a female in her early twenties, Tokyo dialect speaker, gentle tone of voice,'' which was consistent throughout all dialogue lines. We first picked out several dialogue situations from the FKC corpus~\cite{mitsuzawa16} to create realistic dialogue scenarios. The FKC corpus includes approximately 5M text data describing anonymous users' complaints about a specific service or product, such as ``The laundry detergent powder leaves white residue on my clothes.'' Some data optionally contain a user's proposal to solve the issue, e.g., ``I wish the laundry detergent would dissolve more easily,'' and user profile information consisting of age, gender, job, and locale. We chose 10 complaint data containing both the proposal and user profile information for each of eight categories (e.g., ``daily commodities,'' ``food and drink,'' and ``health and beauty'') and created 2 (male or female customer) $\times$ 8 (categories) $\times$ 10 (complaints) $=$ 160 dialogue situations.

{\bf Positive attentive listening:} The second subset considered situations where the operator talks with customers about positive content. For example, a customer feels satisfied with the quality of service, successful transaction, or the employee's behavior and tells the operator about their satisfaction. This design aims to make a voice agent acquire the attentive listening skills essential for social activities in Japanese speech communication~\cite{cook99}. We used the same persona for the operator as that for the first subset and did not specify the personas of customers.

Our CALLS corpus does not include the customers' initial emotions before the dialogues start. The reason is that customers' initial emotions can be roughly classified as negative (angry or sad) or positive (happy), depending on the subset. In addition, our corpus does not contain any angry utterances by the operator because they can make a customer dissatisfied and even sympathize with the operator's angry attitude.

\vspace{-3pt}
\subsection{Crowdsourcing phone-call dialogue lines}\label{subsect:script_collection}
\vspace{-3pt}
We crowdsourced lines of phone calls between the operator and customers. We collected the dialogue lines by microtask crowdsourcing~\cite{simperl15}, where any worker can participate in the task without approval from the client.

For collecting the complaint handling lines, we first prepared Google Forms containing task description (i.e., writing lines of the customer-operator dialogues), personas for the speakers, dialogue turns (4--10), and the dialogue situations. Then, we asked crowdworkers to access the forms and write the dialogue lines. In writing dialogue lines, we instructed crowdworkers to 1) use the user's proposal and make the dialogue as constructive as possible, 2) write each dialogue line following the format ``[speaker's name] [content] [emotion],'' and 3) anonymize the name of a particular company or product if the dialogue situation included it. Finally, we excluded outcomes from spamming workers (e.g., only writing a single character) and workers who did not follow our instructions. 

Microtask crowdsourcing for collecting short (4 turns) attentive-listening lines was similar to that used in Saito et al.'s methodology~\cite{saito22studies}. The instruction to crowdworkers was: ``\textit{Create 4-turn dialogue lines where a customer and operator are talking happily on a phone call. You are free to think about how to end the dialogue, but please consider the assumption that the customer feels happy.}'' We prohibited crowdworkers from including the name of a specific service or product in the dialogue lines. Finally, we proofread all the collected lines and removed ones that 1) never contained positive emotions and 2) included offensive expressions.

\vspace{-3pt}
\subsection{Voice recording}
\vspace{-3pt}
We employed the same female speaker as the STUDIES teacher and asked her to perform the operator's role. She uttered the phone-call lines in a recording studio as if she responded to the customers considering their emotions. The recording was conducted on separate days using a unidirectional
desktop condenser microphone. The sampling rate was 48~kHz.

Although we collected the customers' utterances, we did not record their voices. The reason came from Nishimura et al.'s study~\cite{nishimura22}, which suggests that using only text-based chat history is sufficient to improve speech quality in EDSS.

\begin{table}[t]
\centering
\caption{Number of utterances per each subset. ``CH'' and ``AL'' in this table denote complaint handling and attentive listening, respectively}
\vspace{-3mm}
\label{tab:utterances_per_subset}
\begin{tabular}{c|cc|c}
\hline
\hline
Speaker & CH & AL & Total (hours) \\
\hline
Operator & 2,072 & 1,200 & 3,272 (6.5) \\
Customer & 2,112 & 1,200 & 3,312 (N/A) \\
\hline
\hline
\end{tabular}
\end{table}

\begin{table}[t]
\centering
\caption{Number of utterances for each emotion}
\vspace{-3mm}
\label{tab:utterances_per_emotion}
\begin{tabular}{c|cccc}
\hline
\hline
Speaker & Neutral & Happy & Sad & Angry \\
\hline
Operator & 657 & 1,669 & 946 & 0 \\
Customer & 1,149 & 934 & 284 & 945 \\
\hline
\hline
\end{tabular}
\end{table}

\begin{table*}[t]
\centering
\caption{List of existing Japanese dialogue speech corpora related to CALLS}
\vspace{-3mm}
\label{tab:corpora}
\small
\begin{tabular}{cccccc}
\hline
\hline
Corpus & Dialogue type & Open-sourced & Dur~[hour] & \# of speakers & Emotion label \\
\hline
 Hiraoka et al.~\cite{hiraoka16} & Persuasive & No & 5.7 & 22 & No \\
 Kawahara et al.~\cite{kawahara15} & Attentive listening & No & 2.3 & 8 & No \\
 STUDIES & Empathetic (teacher--student) & Yes & 8.0 & 3 & Yes \\
 \textbf{CALLS (ours)} & Empathetic (operator--customer) & Yes & 6.5 & 1 & Yes \\
\hline
\hline
\end{tabular}
\end{table*}

\vspace{-5pt}
\section{Corpus Analysis}
\vspace{-3pt}
\subsection{Crowdsourcing setting and results}
We used \href{https://www.lancers.jp/}{Lancers} as the crowdsourcing platform for collecting dialogue lines. The overall period of the crowdsourcing ran from July to August 2022. We employed 150 and 1,400 crowdworkers (with overlap per task) to collect dialogue lines for the complaints handling and attentive listening subsets, respectively. We paid crowdworker in the first and second tasks \$4.04 and \$0.4 as compensation, respectively. As a result of our screening, we obtained 820 lines of complaint handling and 600 lines of attentive listening.

\vspace{-3pt}
\subsection{Corpus specification}\label{subsect:spec}
\vspace{-3pt}
Table~\ref{tab:utterances_per_subset} shows the number of utterances per subset of the CALLS corpus consisting of 3,272 utterances by the operator (6.5 hours) and 3,312 utterances by the customers (without their speech). Table~\ref{tab:utterances_per_emotion} shows the number of utterances per emotion label. As we explained in Section~\ref{subsect:scenario}, our corpus does not include any angry utterances by the operator. In addition, the operator and customer utterances are generally imbalanced in terms of emotion labels, which may make modeling the operator's empathetic speaking styles using the interlocutor's emotion more complicated than that for the STUDIES corpus.

\vspace{-3pt}
\subsection{Comparison with existing corpora}
\vspace{-3pt}
Table~\ref{tab:corpora} lists existing Japanese dialogue speech corpora related to the CALLS corpus. Our corpus extends the application range of EDSS technologies suitable for more polite dialogue situations than STUDIES, although ours only includes a single speaker's simulated spoken dialogue lines. However, mixing the CALLS operator's data with the STUDIES teacher's data (about 5 hours) yields the speaker's expressive speech corpus of over 10 hours, comparable to the JSUT corpus~\cite{takamichi20ast} widely used for training high-quality end-to-end TTS models in Japanese.

Our CALLS corpus closely relates to a persuasive dialogue corpus developed by Hiraoka et al.~\cite{hiraoka16}. This dialogue aims to persuade an interlocutor to do something profitable, such as encouraging daily exercises and purchasing a new product. One can regard this dialogue as an example of empathetic dialogues in the task-oriented situation.

The attentive listening subset in our corpus is inspired by a counseling dialogue corpus developed by Kawahara et al.~\cite{kawahara15}. It includes simulated counseling dialogues between six subjects and professional counselors, which contain some acknowledging acts such as ``\jp{うん うん} (un un)'' to show the listener understands and agrees with the speaker~\cite{white89}. Since our CALLS corpus does not cover such behaviors that occur in spontaneous conversation, we will continue to record empathetic spoken dialogues in face-to-face situations.

\begin{figure}[t]
  \centering
  \includegraphics[width=0.9\linewidth, clip]{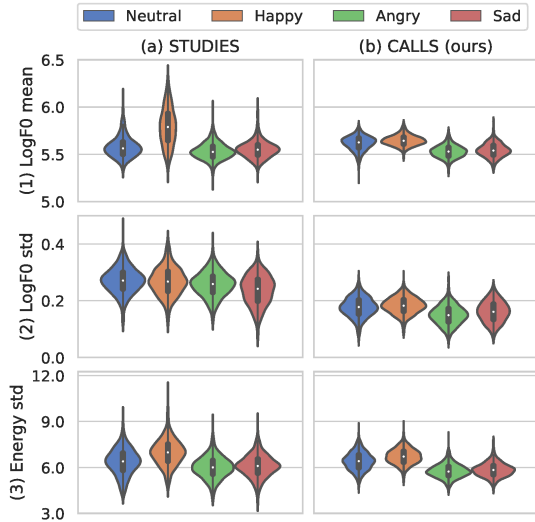}
  \vspace{-8pt}
  \caption{Prosodic feature statistics aggregated based on the interlocutor's emotion.}
  \label{fig:prosody_stats}
\end{figure}

\begin{figure}[t]
  \centering
  \includegraphics[width=0.55\linewidth, clip]{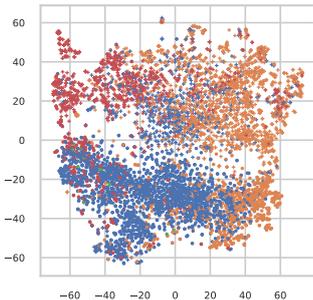}
  \vspace{-8pt}
  \caption{t-SNE plot of sentence embeddings. Colors showing emotion labels correspond to those used in Fig.~\ref{fig:prosody_stats}. Circle and cross marks denote utterances by the STUDIES teacher and CALLS operator, respectively.}
  \label{fig:emb}
\end{figure}

\vspace{-3pt}
\subsection{Domain differences in empathetic spoken dialogues}\label{subsect:analysis_of_domain_differences}
\vspace{-3pt}
We analyzed the differences between our CALLS corpus and existing STUDIES corpus to investigate the effects of domain difference in empathetic spoken dialogues. 

{\bf Prosodic feature differences: } Figure~\ref{fig:prosody_stats} shows the violin plot of the prosodic feature statistics, aggregated on the basis of the interlocutor's emotion. Rows in this figure correspond to the (1) mean and (2) standard deviation (std) of log $F_0$, and (3) std of energy, respectively. We used the WORLD vocoder~\cite{morise16world,morise16d4c} for the $F_0$ extraction. From the results, we can see the the most significant difference is observed in the log $F_0$ mean of voices spoken to the happy interlocutors (Fig.~\ref{fig:prosody_stats}(1)). Concretely, the log $F_0$ mean values for the teacher's voices widely distribute around its median (5.79), while those for the operator's voices narrowly distribute around its median (5.64). Another observation is that the log $F_0$ std values of the teacher's voices are generally higher than those of the operator's voices (Fig.~\ref{fig:prosody_stats}(2)). A similar tendency is observed in the differences of energy std shown in Fig.~\ref{fig:prosody_stats}(3). These results suggest that the expressiveness of speech in empathetic dialogues tends to become greater in casual situations and smaller in formal, polite situations.

{\bf Textual feature differences: } Figure~\ref{fig:emb} shows the t-SNE plot of sentence embedding vectors extracted from the text data of STUDIES teacher and CALLS operator by using BERT~\cite{devlin19}. We used sentence BERT pretrained with Japanese text data available at \href{https://huggingface.co/koheiduck/bert-japanese-finetuned-sentiment}{here}. From this figure, we can see that the embeddings from the two different speakers are clearly isolated on the basis of the values of the $y$-axis, although the values of the $x$-axis roughly differentiate the speaker's emotion regardless of the domain difference. This result shows that texts included in our CALLS corpus cover different ranges from the existing STUDIES corpus.

\vspace{-5pt}
\begin{table}[t]
\centering
\caption{Number of utterances in training/validation/evaluation sets of CALLS operator and STUDIES teacher}
\label{tab:data_split}
\vspace{-3mm}
%\footnotesize
\begin{tabular}{c|rr}
\hline
\hline
%\rule[0mm]{0mm}{2mm}
Set & CALLS & STUDIES \\
\hline
%\rule[0mm]{0mm}{2mm}
Training & 2,824 & 2,209 \\
Validation & 224 & 221 \\
Evaluation & 224 & 211 \\
\hline
\hline
\end{tabular}
\end{table}

\vspace{-5pt}
\section{EDSS Experiment}
\vspace{-3pt}
We report the results of two EDSS experiments: single-/multi-domain EDSS using the CALLS and STUDIES corpora.
\vspace{-3pt}
\subsection{Experimental conditions}
\vspace{-3pt}
This section describes conditions common between the single-/multi-domain EDSS experiments.

{\bf Datasets:} We split speech data of the CALLS operator and STUDIES teacher for training, validation, and evaluation sets as shown in Table~\ref{tab:data_split}. The numbers of utterances per dialogue subset (i.e., complaint handling and attentive listening in CALLS and long-/short-dialogues in STUDIES) were roughly balanced. We downsampled the speech data to 22,050~Hz. We used the validation data to choose hyperparameters for our backbone TTS model, whose parameters were randomly initialized.

{\bf Backbone TTS model:} An acoustic model for predicting a mel-spectrogram from text was FastSpeech 2 (FS2)~\cite{ren21}, with the PyTorch implementation for Japanese TTS provided at \href{https://github.com/Wataru-Nakata/FastSpeech2-JSUT}{here}. We followed the settings of a neural network architecture and speech parameter extraction of this implementation. The optimizer was Adam~\cite{kingma14adam} with a learning rate $\eta$ of 0.0625, $\beta_1$ of 0.9, and $\beta_2$ of 0.98. We first pretrained FS2 by using the JSUT corpus~\cite{takamichi20ast} with 200K iterations and then fine-tuned it by using the training data with 100K iterations. A neural vocoder for synthesizing speech from mel-spectrogram was HiFi-GAN~\cite{kong20}, with \href{https://github.com/jik876/hifi-gan}{the official PyTorch implementation}. We trained HiFi-GAN by using the training data with 350K iterations. The optimizer was Adam with $\eta$ of 0.0002, $\beta_1$ of 0.8, and $\beta_2$ of 0.99. 

{\bf Evaluation criteria:} We conducted two kinds of five-scaled mean opinion score (MOS) tests regarding the naturalness and speaking-style similarity of synthetic speech. In the naturalness test, we presented 30 synthetic speech samples to listeners in random order. Listeners rated the naturalness of each speech sample. In the similarity test, listeners first listened to HiFi-GAN-vocoded natural speech as reference. Then, listeners scored how similar the speaking style of the presented voice was to the reference. The number of presented synthetic speech samples was 30. We conducted 8 MOS tests to evaluate the naturalness or similarity of speech samples by the CALLS operator or STUDIES teacher synthesized with single-/multi-domain EDSS model. Fifty listeners participated in each test using our crowdsourcing evaluation system, and the total number of listeners was 400.

\vspace{-3pt}
\subsection{Evaluation of single-domain EDSS}
\vspace{-3pt}
We independently trained two EDSS models using the CALLS or STUDIES corpus, which made each model focus on the given dialogue domain only. Following Saito et al.'s work~\cite{saito22studies}, we investigated three conditional features: 1) the agent's emotion (\textbf{A-Emo)}, 2) the user's emotion (\textbf{U-Emo}), and 3) context information learned from a conversational context encoder (\textbf{CCE})~\cite{guo21}. The \textbf{CCE} extracted an embedding vector of dialogue context from joint vectors of one-hot coded speaker identity and sentence embedding sequences (one current and three previous utterances) obtained by using the same BERT as we explained in Section~\ref{subsect:analysis_of_domain_differences}. The dimensionality of BERT embedding was 768, and we projected the embedding onto the 256-dimensional feature space using a fully-connected layer. 

Figure~\ref{fig:mos}(1) shows the evaluation results. For the two models trained only on the CALLS or STUDIES corpus, the highest naturalness MOS values were different (3.82 and 3.59, respectively). In addition, the latter was generally inferior to the former regarding the similarity MOS values. These results indicate that the CALLS operator's speaking styles are easier to reproduce than the STUDIES teacher's because the former have lower variances of prosodic features. Among the three compared factors, \textbf{CCE} improved both the naturalness and similarity, and the combination of \textbf{CCE} and \textbf{A-Emo} achieved the highest similarity MOS for the CALLS operator, which suggest that \textbf{CCE} by itself can learn expressive speaking styles and additional conditioning by emotion labels may enhance this ability.

\vspace{-3pt}
\subsection{Evaluation of multi-domain EDSS}
\vspace{-3pt}
We trained a single EDSS model using the mixture of CALLS and STUDIES corpora, which increased the number of training data but made the training more challenging due to the domain differences. Here, we set the baseline model as \textbf{FS2\&CCE} in accordance with the results of the previous experiment and considered conditioning the model by a speaker embedding (\textbf{SPK}) as an additional factor to \textbf{A-Emo} and \textbf{U-Emo}. This investigation aimed to clarify whether we should explicitly separate the different empathetic speaking styles by the two speakers during the EDSS model training or not.

\begin{figure}[t]
  \centering
  \includegraphics[width=0.99\linewidth, clip]{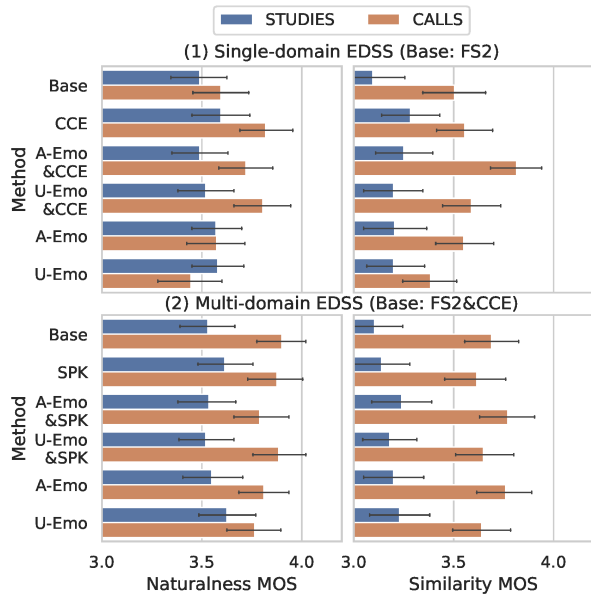}
  \vspace{-8pt}
  \caption{MOS test results with their 95\% confidence intervals}
  \label{fig:mos}
\end{figure}

Figure~\ref{fig:mos}(2) shows the evaluation results. From the results, learning with multiple empathetic dialogue speech corpora improved the overall naturalness and style similarity of the CALLS operator's synthetic speech. However, the improvement for the STUDIES teacher's voices was limited and the achievable style similarity was less than when trained on a single corpus. These results suggest that, even if we condition an EDSS model by speaker and emotion labels, we need to incorporate a method, such as the domain-adversarial training~\cite{ganin16}, for capturing the characteristics of different domains explicitly.

\vspace{-5pt}
\section{Conclusion}
\vspace{-3pt}
We presented CALLS, a new speech corpus to advance multi-domain empathetic dialogue speech synthesis (EDSS) research. Our corpus includes the same female speaker as the teacher in the STUDIES corpus, acting as an operator in simulated phone calls. This design enables a single speaker's different empathetic speaking styles to be analyzed in multiple dialogue domains. The results of EDSS experiments showed that mixing the two corpora during the training caused biased improvements in the quality of synthetic speech due to the different degrees of expressiveness. In future, we introduce data augmentation~\cite{terashima22} to cope with the domain differences in multi-domain EDSS and evaluate our empathetic spoken dialogue system in face-to-face communication.

\textbf{Acknowledgements:}
This research was conducted as joint research between LINE Corporation and Saruwatari-Takamichi Laboratory of The University of Tokyo, Japan. Part of this work was supported by JSPS KAKENHI Grant Number 21K21305.

   %\newpage
   \ninept
   \bibliographystyle{IEEEtran}
   \bibliography{template.bbl}

\end{document}